# Working Mechanism and Behavior of Collison Nebulizer

James Q. Feng*, Liang-Sin Go, Jenny Calubayan, Robert Tomaska

Optomec, Inc.,2575 University Avenue, #135, St. Paul, Minnesota, USA

*Corresponding author: jfeng@optomec.com

**Abstract**

The Collison nebulizer (as well as its variations) has been widely used for generating fine aerosol droplets of a few microns from liquids of viscosity up to 1000 centipoise. It was originally developed for producing medical aerosols in inhalation therapy, and now has become an important component as pneumatic atomizer in the Aerosol Jet® direct-write system for additive manufacturing. Qualitative descriptions of its working mechanism can be found in the literature as an expanding high-speed gas jet creates a negative pressure to syphon liquid into the jet stream, where the liquid is subsequently blown into sheets, filaments, and eventually droplets. Yet quantitative analysis and in-depth understanding have been lacking until rather recently. In this work, we present a logical discussion of the working mechanism of Collison nebulizer based on OpenFOAM® CFD analysis of compressible jet flow in the jet expansion channel. The positive-feedback liquid aspiration mechanism becomes clear by examining the CFD results as the jet expansion channel geometry is varied. As a consequence, the output mist density can be rather insensitive to the liquid viscosity, which is illustrated by a set of experiments with the Collison-type pneumatic atomizer in an Aerosol Jet® direct-write system. Thus, an intrinsic self-regulation mechanism is elegantly incorporated in the Collison nebulizer design. As intuitively expected, experimental data also supports the notion of the existence of an upper limit for liquid holdup in the limited space of jet expansion channel; therefore, the output mist density cannot increase indefinitely by increasing the atomization gas flow rate.

## 1. Introduction

Many technological applications involve aerosolizing liquid materials into fine mist droplets as the key process enabler. Therefore, various nebulizers have been developed to generate microdroplets—droplets of a few microns in diameter—that are required primarily for the treatment of respiratory diseases via medical aerosol inhalation therapy (May 1973; Hess 2000). An outstanding example is the Collison nebulizer introduced by W. E. Collison for the inhalation therapy (at a meeting of the British Medical Association in 1932, and in the book of Collison 1935), which has later been extended to other applications such as Aerosol Jet® printing for additive manufacturing (Feng and Renn 2019). It is the important role played by the Collison nebulizer in Aerosol Jet® printing that motivated us to carry out an in-depth study of its working mechanism and behaviour.

As shown in Fig. 1, the Collison nebulizer uses a high-speed jet of compressed air through a small orifice to create a reduced (or 'negative') pressure in the jet expansion channel that can syphon liquid from the reservoir into the jet stream, and to blow the liquid into thin sheets, filaments, and eventually droplets. Like typical two-fluid "air-blast" atomization process, the liquid droplets so generated have a wide range of sizes. Most of those droplets are too large for effective inhalation therapy, where the most desirable droplets are those with diameters of 1 to 5 microns, because larger droplets could not reach deep lung channels due to impaction and sedimentation losses whereas smaller ones could not be deposited onto lung channel walls due to lack of sufficient inertia (Niven and Brain 1994; Hess 2000). To sift out the microdroplets for the mist output, those atomized liquid droplets are carried by the jet flow toward the wall of nebulizer chamber, where large droplets with sufficient mass are deposited by inertial impaction. Consequently, only a small fraction (typically < 0.1%) of the liquid syphoned into jet stream can become fine enough droplets (e.g., < 5 μm) to escape the wall impaction and flow out as the output mist (May 1973).

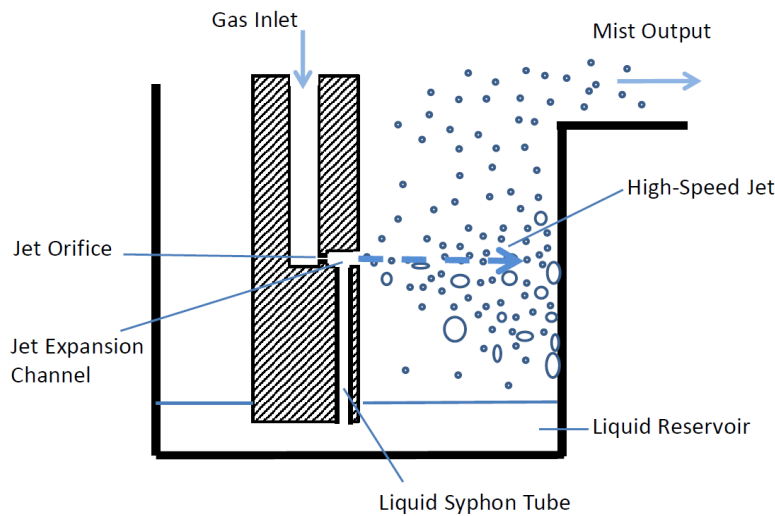

Fig. **1**. Schematic of a typical configuration of the Collison nebulizer

Because more than 99.9% of the liquid going through the atomization process goes back to the liquid reservoir, the liquid aspiration rate through the liquid syphon tube could substantially influence the nebulizer output mist density. Adequate liquid aspiration rate requires adequate negative pressure in the jet expansion channel. To obtain a given aspiration rate, a liquid with higher viscosity would require proportionally stronger negative pressure as alluded to by Poiseuille's law for channel flow. The empirical knowledge gained over years of application with the Aerosol

Jet® printing indicated that the Collison nebulizer could generate reasonably dense mist with inks of viscosities across two orders of magnitude, while the atomization gas flow rate varied only within a factor of 2 (which would be expected to yield a negative pressure variation by about a factor of 4 according to Bernoulli's principle.) A recent CFD analysis based on the OpenFOAM® C++ Toolbox revealed a self-regulating mechanism for liquid aspiration rate control (Feng, 2018a). Hence, the output mist density of the Collison nebulizer can be relatively insensitive to the liquid viscosity, to enable the Aerosol Jet® direct-write system to print a great variety of ink materials for additive manufacturing (Feng and Renn 2019).

Here, the behaviour of Collison nebulizer is discussed in the context of Aerosol Jet® direct-write technology with the liquid material for atomization often called "ink". When used in an additive manufacturing process, the Collison nebulizer (also called "pneumatic atomizer" in the Aerosol Jet® system) offers the attractive attribute of configuration simplicity with self-aspiration of ink material (which eliminates the need of an active pump). It is also capable of generating printable ink mist from a wide range of liquid materials by virtue of its relative insensitivity to the ink rheological properties. There is a great desire to gain in-depth understanding of the working mechanism of this type of pneumatic atomizer. Yet in the literature, the technical discussion about fluid dynamic fundamentals of Collison nebulizer performance remained scarce, despite its wide usage in a variety of applications. The only noticeable paper is that published by May (1973), providing some details of the geometric dimensions and experimental data from scientific measurements. Majority of other authors (e.g., Niven and Brain 1994; Hess 2000) tend to focus on experimental tests for comparing functional performance of various pneumatic (or "air-jet") nebulizers rather than enhancing scientific understanding of the complicated fluid dynamic process above the usual qualitative handwaving.

In what follows, we begin by an analysis of the basic mechanism for liquid aspiration in the Collison nebulizer in view of our OpenFOAM® CFD results (Feng 2018a). Then, we discuss some general behaviour of the Collison-type pneumatic atomizer in Aerosol Jet® printing for additive manufacturing with experimental measurements. Although our discussion is mostly presented in the context of Aerosol Jet® printing, the enhanced fundamental understanding should be beneficial to general applications of the Collison nebulizer. The conclusions are provided at the end of this paper.

## 2. Mechanism for Liquid Aspiration

### 2.1. Results of the CFD Analysis

To bring liquid ink from the reservoir though liquid syphon tube up to the jet expansion channel for atomization, the magnitude of negative pressure in the jet expansion channel must be sufficient to overcome the hydrostatic pressure due to downward pointing gravity. In the standard Collison nebulizer configuration, the jet expansion channel is located about $h = 20$ mm above the liquid level in the reservoir (cf. Fig. 1). Thus we expect to have a hydrostatic pressure of $\rho_{ink}\ g\ h \sim 400$ Pa if the ink density $\rho_{ink}$ is taken as 2 g/cc, $g$ = 9.81 m s$^{-2}$ and $h$ = 0.02 m. Then, the atomization of ink should occur when the gas jet flow rate is above a threshold value that induces a negative pressure with magnitude $\Delta P$ > 400 Pa.

The OpenFOAM® CFD analysis (Feng 2018a) is based on an axisymmetric model of compressible gas flowing from an inlet channel through a small jet orifice into a jet expansion channel of larger diameter and then into a much larger atomization chamber with a solid wall at its end. The steady-state numerical solutions were computed with a compressible flow solver called *rhoSimpleFoam* with a *k-ε* turbulence model in the Reynolds Averaged Navier-Stokes equations (which includes equations for conservations of mass, momentum, and energy, governing the flow of a fluid described by the ideal gas law and Fourier's law of heat conduction with Sutherland's law for dynamic viscosity, along with equations of a common *k-ε* model to account for the turbulence effects). As a nominal model setting, the jet orifice has a diameter of $D$ = 0.35 mm and the diameter and length of jet expansion channel are $D_e$ = 1.5 mm and $L_e$ = 2.7 mm, to be consistent with the standard Collison nebulizer design (May 1973).

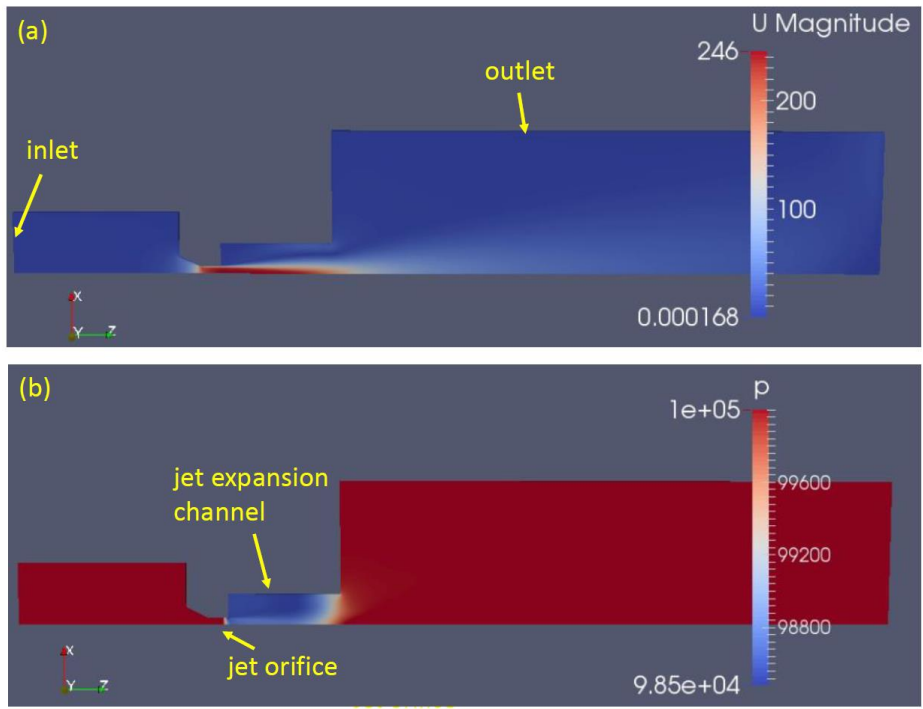

Fig. 2. OpenFOAM® CFD results: (a) The field of gas velocity magnitude $|U|$ (m/s), and (b) pressure $p$ (Pa) for the nominal case configuration at gas flow rate of $Q$ = 1200 sccm

Fig. 2 shows that the jet velocity can approach 246 m/s (corresponding to a Mach number $Ma$ = 0.746) at the exit of the jet orifice when the gas flow rate $Q$ = 1200 sccm (where "sccm" stands for the "standard cubic centimeters per minute"). Then, the jet expands with velocity decreasing as it moves forward. A significant region of negative pressure with $\Delta P \sim 1524$ Pa indeed appears in the jet expansion channel, providing the syphoning effect for liquid aspiration. Noteworthy here is that the pressure field does not exhibit the similar structure as that of the flow velocity; the lowest pressure does not coincide with the highest velocity as intuitively anticipated from Bernoulli's principle. This phenomenon can be explained by a detailed examination of the gas flow field in the jet expansion channel where a back flow exists near the channel wall surrounding the jet, by virtue of mass conservation (Feng 2018a). Due to viscous drag, the jet flow tends to bring more gas out of the jet expansion channel than what is supplied from the jet orifice, which creates a reduced local pressure to drive the back flow for compensating the jet depleted gas. As a consequence, a region of negative pressure must appear in the jet expansion channel.

Using the CFD model, the effect of gas flow rate on the negative pressure magnitude $\Delta P$ can be conveniently examined. For the nominal case ($D = 0.35$ mm, $D_e = 1.5$ mm and $L_e = 2.7$ mm), the values of computed $\Delta P$ (Pa) for various gas flow rate $Q$ (sccm) can be fitted into a linear equation (for $600 \leq Q \leq 1800$),

$$\Delta P = 2.0367 \, Q - 896.6 \quad \text{(with } R^2 = 0.9966) \, , \tag{1}$$

which indicates $\Delta P = 427$ Pa at $Q = 650$ sccm and $\Delta P = 2769$ Pa at $Q = 1800$ sccm. Thus, the value of $\Delta P$ predicted by (1) suggests that for $\Delta P > 400$ Pa (to overcome the hydrostatic pressure of ink in the liquid syphon tube as discussed above) we at least need a flow rate $Q > 600$ sccm, which is consistent with our empirical knowledge that observable mist output from the Aerosol Jet® pneumatic atomizer usually occurs with $Q > 700$ sccm (as illustrated in Fig. 6) and adequate mist output for practical Aerosol Jet® printing can often be obtained at $Q \sim 1200$ sccm (for inks of density $\rho_{ink} \sim 2$ g/cc).

Furthermore, the computed gauge pressure upstream to the jet orifice $P_g$ ($= p_{max} - 10^5$ Pa where $10^5$ Pa is assumed to be the ambient pressure in the atomization chamber) and jet Mach number $Ma$ (for $600 \leq Q \leq 1800$) can be expressed as

$$P_g = 0.0256 \, Q^2 + 6.9095 \, Q - 2500$$

and $$\tag{2}$$

$$Ma = 0.0005 \, Q + 0.0836 \quad \text{(with } R^2 > 0.998.) \, ,$$

indicating that $P_g = 0.128$ and $0.929$ bar (where 1 bar = $10^5$ Pa), $Ma = 0.409$ and $0.984$ for $Q = 650$ and 1800 sccm. The CFD results for $P_g$ agree reasonably with our laboratory gauge pressure measurements of around 0.12 and 0.85 bar for $Q = 650$ and 1800 sccm (despite the simplifications and idealizations assumed in the CFD model).

By varying jet expansion channel parameters from those for the nominal case, the effect of channel geometry can be investigated. For example, if the diameter of jet expansion channel $D_e$ is reduced to 1.0 mm (from the nominal 1.5 mm), the fitted equation for $\Delta P$ versus $Q$ becomes

$$\Delta P = 0.0021\, Q^2 + 2.2406\, Q - 767 \text{ (with } R^2 = 0.9999)\,, \tag{3}$$

which indicates $\Delta P = 1576$ Pa at $Q = 650$ sccm and $\Delta P = 10070$ Pa at $Q = 1800$ sccm, much greater than that for the nominal case as given in (1). The CFD results given by (3) for $D_e = 1.0$ mm suggests a significant enhancement of $\Delta P$ by reducing the jet expansion channel diameter. As expected, the computed results also show a significant reduction of $\Delta P$ by increasing $D_e$ to 1.7 mm (from 1.5 mm), e.g., $\Delta P = 828$ Pa at $Q = 1200$ sccm whereas (1) for $D_e = 1.5$ mm indicates $\Delta P = 1547$ Pa at $Q = 1200$ sccm. Moreover, the CFD results also reveal the fact that increasing the jet expansion channel length $L_e$ can enhance $\Delta P$, consistent with the back flow mechanism in the jet expansion channel. This is because to sustain a given amount of back flow in a longer and narrower channel invokes a greater pressure drop (Feng 2018a).

## 2.2. Implications Revealed by the CFD Results

Interestingly, the findings with OpenFOAM® CFD analysis reveal a positive-feedback mechanism for liquid aspiration in the Collison nebulizer, depicted in Fig. 3. At a low gas flow rate (e.g., $Q \sim 650$ sccm or so), the magnitude of negative pressure $\Delta P$ may barely be enough to bring the liquid ink up to the jet expansion channel with a very low aspiration rate. As the liquid accumulates in the channel, the effective channel diameter for gas flow shrinks and subsequently $\Delta P$ would increase, leading to greater liquid aspiration rate until a dynamic balance of the liquid holdup is established with the liquid removal rate by the blasting jet flow. With the increasing amount of liquid in the channel, the blasting gas velocity is also expected to increase until reaching a dynamic balance. As the liquid ink is blown out of the channel (in an air-blast atomization process), it forms liquid sheets, filaments, and eventually droplets when sheared by the high-speed jet flow.

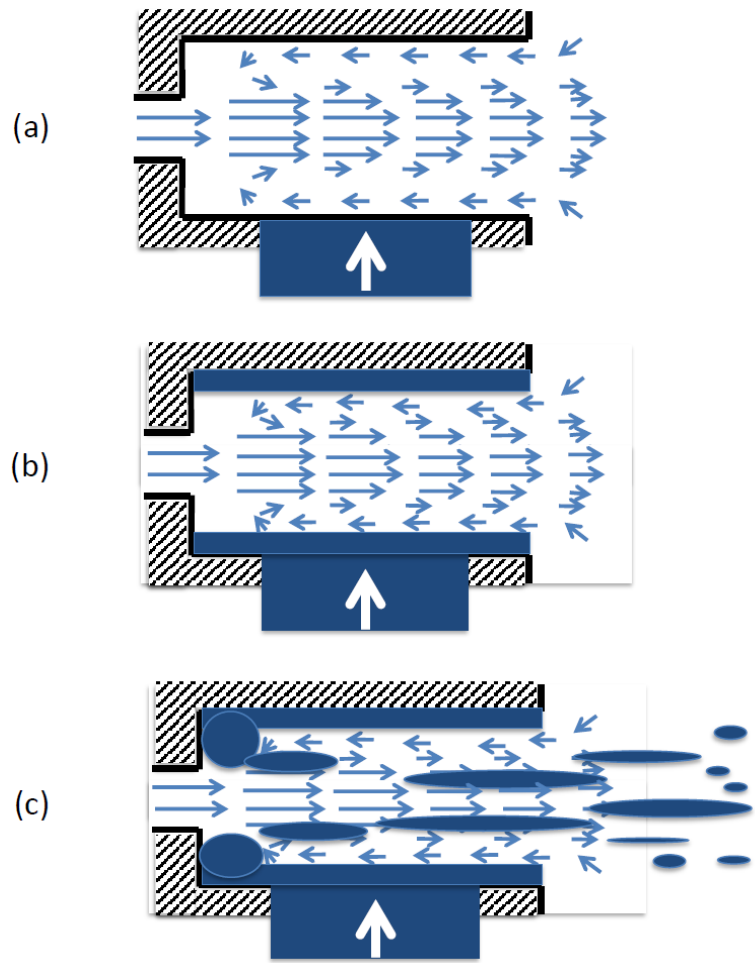

Fig. 3. Fluid dynamic process in the jet expansion channel: (a) Liquid is syphoned into jet expansion channel when $\Delta P$ exceeds the hydrostatic pressure, (b) Liquid accumulates in the channel effectively reducing the channel diameter $D$ (for gas flow), and thereby enhancing $\Delta P$ to increase liquid aspiration rate, and (c) Sustained continuous atomization occurs with appropriate ratio of supplied liquid and gas flow rates.

Some early experimental measurements of a Collison nebulizer (May 1973) showed liquid aspiration rate about 67 cc/min (per jet) for water. This requires an extra pressure difference of about 180 Pa over the syphon tube with a length of 20 mm and diameter of 1.5 mm, assuming a liquid viscosity of 1.0 cp based on Poiseuille's law, in addition to the 200 Pa of hydrostatic pressure (for a liquid density of 1,0 g/cc). Hence, a negative pressure of magnitude $\Delta P = 380$ Pa generated with a gas flow rate of $Q \sim 600$ sccm or with an "air pressure" $P_g \sim 0.109$ bar, according to (1) and (2), should be sufficient. But in reality, the gas flow rate used with the Collison nebulizer was typically $Q > 2000$ sccm (May 1973), with supersonic jet ($Ma > 1$) and $\Delta P > 3000$ Pa according to equations (1) and (2). This fact suggests that an upper limit for liquid holdup also exists in the finite space of jet expansion channel and as liquid in the channel increases beyond an optimal amount, the suction force for liquid aspiration diminishes. Thus, without active liquid pump, the Collison nebulizer produces a self-regulating mechanism for liquid aspiration in its atomization process.

### 3. Pneumatic Atomizer in Aerosol Jet® System

3.1. Aerosol Jet® Printing with Pneumatic Atomizer

As an additive manufacturing equipment, the Aerosol Jet® system deposits ink material in a form of high-speed collimated mist stream of microdroplets (Feng and Renn 2019), as shown in Fig. 4. The diameters of those microdroplets, (typically in the range of 1 to 5 μm) happen to be similar to those used in the inhalation therapy for the similar reasons. Thus, the output mist from the Collison nebulizer, once appropriately conditioned, can be used for Aerosol Jet® printing, which has enabled various applications such as printed electronics on substrate of complex shapes, fabrication of antennas and sensors that conform to the 3D shape of the underlying substrate, etc. In fact, the Collison nebulizer can be used in an Aerosol Jet® system directly as the atomizer in Fig. 4 (with a companion virtual impactor described as follows).

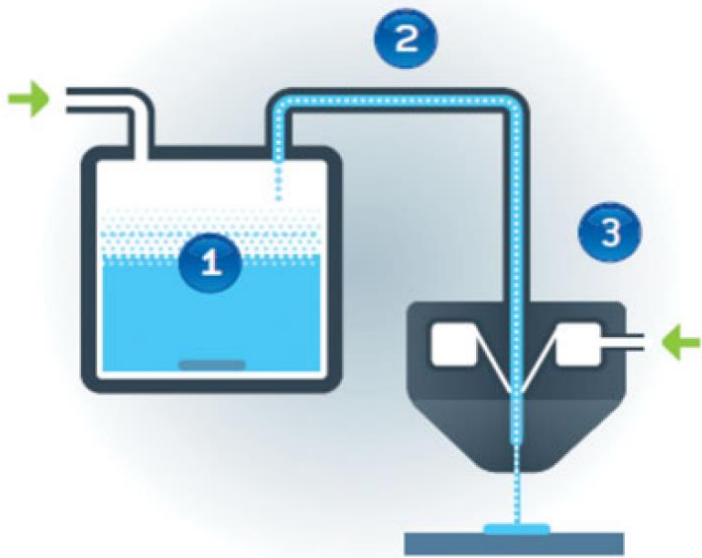

Fig. 4. Aerosol Jet® direct-system system: (1) atomizer that generates mist of microdroplets of functional inks; (2) mist transport-conditioning channel that delivers a concentrated mist of ink microdroplets; (3) deposition head to form high-speed collimated mist stream through an aerodynamic focusing nozzle with sheath gas (with arrows indicating the mist carrier gas inlet and nozzle sheath gas inlet).

However, the typical mist flow rate (through a single ink deposition nozzle) for Aerosol Jet® printing is often less than 500 sccm (depending on print feature size), whereas the Collison-type pneumatic atomizer outputs a mist at a much higher flow rate due to the requirement for atomization (i.e., typically with $Q > 1000$ sccm for producing adequate mist density).    To convert mist flow rate from the pneumatic atomizer output to that for Aerosol Jet[®] deposition, a virtual impactor (cf. Marple and Chien 1980; Loo and Cork 1988) is used for mist flow conditioning with the minor flow through its collector being delivered to the Aerosol Jet[®] deposition nozzle ( [https://www.optomec.com/printed-electronics/aerosol-jet-technology](https://www.optomec.com/printed-electronics/aerosol-jet-technology) ). Because limited flow rate can be used for a given deposition nozzle size (Feng, 2019; Feng and Renn, 2019), high concentration of ink microdroplets in the mist flow (in terms of "mist density") is desired for high-throughput Aerosol Jet[®] printing.    While removing the excess gas in the pneumatic atomizer mist output, the virtual impactor has been found to offer an extra benefit to increase mist density (as illustrated in Fig. 5) often by a factor of 2 to 5.

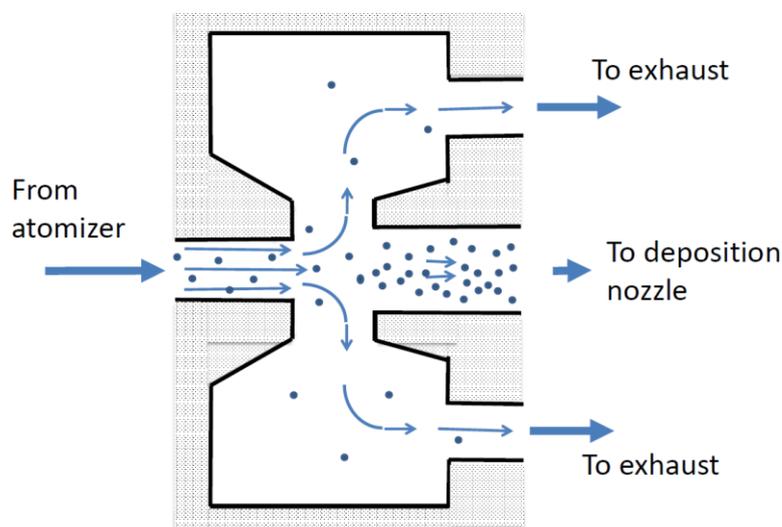

Fig. 5. Mist flows in a virtual impactor: majority of mist droplets from atomizer could not follow the sharply bending streamlines of the exhaust flow and become concentrated in the "minor flow" toward the deposition nozzle with appropriated flow rate

As intended for separating aerosol particle by their aerodynamic sizes, the virtual impactor has been designed based on the mechanism of inertial impaction. Due to their significant inertial effect, the large particles cannot follow the sharply bending streamlines of the "major" flow and are going straight through the "collector" with the "minor" flow, whereas relatively small particles move with the major flow to an exhaust outlet.    The "cut-off" point that separates large and small particles can be adjustable by the inlet flow rate and the device geometry.

In applications for printed electronics, the pneumatic atomizer ink usually consists metal nanoparticles with polymeric "capping agents" suspended in solvents with a solid fraction >60 wt% and viscosity in a range of 20 to 1000 cp (depending on ink formulation) with a liquid ink density of 2 to 3 g/cc.    One of the important performance indicators is the solid mass output from a deposition nozzle, in units of milligrams per minute (mg/min) as a measure of the print throughput. To remove the mist flow rate dependence, which can vary based on print feature size with different nozzle size, the value of mist density (as the solid mass deposition rate divided by mist flow rate) becomes a more meaningful indicator for evaluating the atomization performance.

## 3.2. Output Mist Density of Aerosol Jet® Pneumatic Atomizer

To examine performance of the Collison-type pneumatic atomizer for Aerosol Jet® printing, we measure mist density in the mist flow at the exit of deposition nozzle for a range of relevant atomizer gas flow rate $Q$. With a given mist output from the pneumatic atomizer, the mist density delivered to the deposition nozzle is known to depend on the mist flow rate which can influence the mist mass loss in mist transport channel due to gravitational wall deposition as well as impaction onto channel wall (Feng 2018b). Therefore, we carry out the experiment using a fixed mist flow rate (e.g., at 75 sccm) to the deposition nozzle. This ensures the observed change in mist density, determined from the deposited ink mass when varying the atomization flow rate $Q$, to directly correlate to the pneumatic atomizer mist output (assuming the virtual impactor performance is relatively insensitive to the variation of atomization flow rate).

For a fixed mist flow rate to the deposition nozzle by varying the atomization gas flow rate $Q$, the mist density (mg/cc) can be determined as a function of the atomization flow rate. Shown in Fig. 6 is the relationship of mist density versus atomization gas flow rate for three different silver nanoparticle inks; one has a viscosity of 30 cp, another about 200 cp, and the other about 1000 cp (as measured with the Brookfield viscometer). Not surprisingly, the threshold gas flow rate for inception of ink atomization seems to be around 600 to 900 sccm. It is quite clear that higher mist density can usually be obtained at a given atomization gas flow rate with inks of lower viscosity, especially for $Q < 1500$ sccm. However, the differences in mist density at a given $Q$ for inks of different viscosity values are much smaller (i.e., within the same order of magnitude) than the viscosity value differences (over orders of magnitude). If we focus on the peak values of mist density, they can be rather comparable despite the ink viscosity varies over almost two orders of magnitude. As often observed with Aerosol Jet® printing, increasing the atomization gas flow rate in the Collison-type pneumatic atomizer does not yield significantly higher mist density beyond $Q = 1300$ sccm, except for the very viscous ink (e.g., that with viscosity ~1000 cp) which reaches the maximized mist density around $Q = 2000$ sccm.

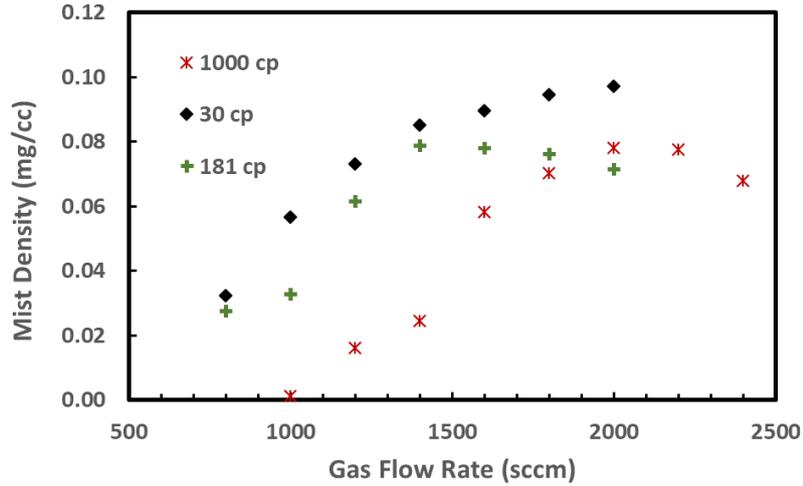

**Fig. 6.** Mist density in the mist flow through Aerosol Jet® deposition nozzle versus atomization gas flow rate for silver nanoparticle inks of different viscosities (1000 cp by DOWA, 30 cp by pvnanocell, and 181 cp by PARU).

From a simplistic theoretical point of view, the relationship among liquid aspiration rate $Q_{ink}$ and liquid viscosity $\mu_{ink}$ as well as the pressure difference $\Delta P$ across the liquid syphon tube length $L_s$ may be approximately described by the Poiseuille equation

$$Q_{ink} = [\pi D_s^4 / (128\, L_s)]\; \Delta P / \mu_{ink} \qquad , \qquad\qquad (4)$$

with $D_s$ denoting the diameter of the liquid syphon tube.  Thus the liquid aspiration rate $Q_{ink}$ through the liquid syphon tube (of fixed $D_s$ and $L_s$) could be maintained as an invariant if $\Delta P$ increases proportionally to the increase of $\mu_{ink}$ .

In the context of Fig. 3, we have discussed the mechanism of $\Delta P$ increasing with the amount of liquid holdup in the jet expansion channel. Physically, the amount of liquid holdup is expected to increase with liquid viscosity because a more viscous liquid would become more difficult for the jet flow to remove liquid out of the jet expansion channel.  Therefore, relatively higher $\Delta P$ should appear when inks of higher viscosity are atomized. Then, the liquid aspiration rate $Q_{ink}$ becomes rather insensitive to the ink viscosity, and this explains the experimental data illustrated in Fig. 6.

However, it should be noted that inks of similar shear viscosity values may not always behave similarly in terms of atomizer output mist density.   Other rheological properties, such as elasticity, extensional viscosity, as well as surface tension, can also significantly alter the atomization behaviour (Lefebvre 1989).   Detailed analysis of liquid atomization process, such as evolution of liquid sheets, filament, and droplets, for inks of different rheological properties, can certainly help gain important insights for further understanding and improving design of the Collison-like pneumatic atomizers. Yet this kind of efforts would be much more complicated and resource demanding than what can be pursued within the present research scope.

Although higher atomization gas flow rate $Q$ tends to generate stronger suction force for ink aspiration as indicated by (1), optimal atomization performance is expected when the gas and liquid flow rates are adjusted to an appropriate ratio (Lefebvre, 1989). But the self-regulated liquid aspiration rate in Collison nebulizer may not always provide such an optimal "gas-to-liquid ratio (GLR)" with different gas flow rates. In fact, experiments of Collison nebulizer output for water atomization at various gas flow rates (May 1973) show decreasing mist density with increasing gas flow rate when $Q > 2000$ sccm (per jet). Thus, maximum mist density is usually obtained with $Q < 2000$ sccm, although the actual atomization outcome can vary significantly with the art of ink formulation.

Fig. 6 also indicates that the pneumatic atomizer can produce a mist density that is rather insensitive to the ink viscosity (which is often an important parameter for atomization although the measured shear viscosity is not the only factor determining the atomization performance). Liquid inks of low viscosity can easily flow through the syphon tube into the jet expansion channel to establish such a dynamic balance between liquid aspiration and removal rates. Higher viscosity inks would require greater $\Delta P$ for a desirable aspiration rate through the syphon tube at a given gas flow rate, but they are also more difficult to remove by the blowing jet flow. Thus, the liquid holdup volume is expected to increase in the jet expansion channel for inks of higher viscosity, yielding a greater $\Delta P$ for adequate aspiration force. This could be the enabling mechanism for the Collison nebulizer to atomize inks of viscosities spanning more than two orders of magnitude with the gas flow rate (as well as the value of $\Delta P$ according to (1)) varying only within a factor of 4 or so (as expected from Bernoulli's principle commented in Introduction).

## 4. Conclusions

In view of the CFD results and available experimental data, the working mechanism and behaviour of the Collison nebulizer can be theoretically described in terms of fluid dynamics. The self-regulating liquid aspiration mechanism becomes clear by examining the CFD results as the jet expansion channel geometry is varied. The pressure gradient for liquid aspiration is shown to increase with reducing the jet expansion channel diameter, which is equivalent to syphoned liquid filling up the channel space. During the atomization process, a dynamic balance of liquid holdup is expected in the jet expansion channel for liquid input by syphoning and liquid removal by blowing gas jet flow. The amount of liquid hold up could increase with the liquid viscosity due to increased difficulty of liquid removal, leading to stronger syphon force as needed to keep up with the liquid aspiration rate. Consequently, the Collison-type pneumatic atomizer can process inks with a wide range of viscosity values for comparable output mist density, which is quite desirable for various Aerosol Jet® applications.

As intuitively expected, experimental data supports the notion of existence of an upper limit for liquid holdup in the limited space of jet expansion channel; therefore, the output mist density cannot increase indefinitely by increasing the atomization gas flow rate. But its mechanical simplicity (i.e., without moving parts and active pump) and insensitivity to liquid rheological properties make the Collison nebulizer an attractive means for atomization in the Aerosol Jet® direct-write system for additive manufacturing.

**References**

Collison WE (1935) Inhalation Therapy Technique. Heinemann, London

Feng JQ (2018a) A computational analysis of gas jet flow effects on liquid aspiration in the Collison nebulizer. Proceedings of the 5th International Conference on Fluid Flow, Heat and Mass Transfer (FFHMT'18), Paper No. 180. https://doi.org/10.11159/ffhmt18.180

Feng JQ (2018b) Multiphase flow analysis of mist transport behavior in Aerosol Jet® system. International Journal of Computational Methods and Experimental Measurements 6(1): 23-34.   https://doi.org/10.2495/CMEM-V6-N1-23-34

Feng JQ (2019) Mist flow visualization for round jets in Aerosol Jet® printing. Aerosol Science and Technology 53(1): 45-52. https://doi.org/10.1080/02786826.2018.1532566

Feng JQ, Renn MJ (2019) Aerosol Jet® direct-write for microscale additive manufacturing. Journal of Micro- and Nano-Manufacturing 7(1): 011004. https://doi.org/10.1115/1.4043595

Hess DR (2000) Nebulizers: principles and performance. Respiratory Care 45(6): 609-622.

Lefebvre AH (1989) Atomization and Sprays. Hemisphere Publishing Corp.

Loo BW, Cork CP (1988) Development of high efficiency virtual impactors. Aerosol Science and Technology 9: 167-176. https://doi.org/10.1080/02786828808959205

Marple VA, Chien CM (1980) Virtual impactors: A theoretical study. Environmental Science and Technology 14(8): 976-985. https://doi.org/10.1021/es60168a019

May KR (1973) The Collison nebulizer: description, performance and application. Journal of Aerosol Science 4(3): 235-243. https://doi.org/10.1016/0021-8502(73)90006-2

Niven RW, Brain JD (1994) Some functional aspects of air-jet nebulizers. Int. J. Parmaceut. 104: 73-85. https://doi.org/10.1016/0378-5173(94)90338-7